\documentclass [12pt,a4paper]{article}
\usepackage{epsfig,amsmath,amssymb}
\usepackage{natbib}
\usepackage{bm}
\usepackage{psfrag}
\usepackage{wrapfig}
\usepackage{booktabs}
\usepackage{geometry}
\usepackage{pdflscape}
\def\e{\hbox{E}}
\def\cov{\hbox{Cov}}

\def\se{\hbox{SE}}

\newcommand{\F}{\mathcal{F}}

\usepackage{xcolor,colortbl}
\definecolor{Gray}{gray}{0.85}
\newcolumntype{a}{>{\columncolor{Gray}}r}

\begin{document}

\title{A Simple and Effective Inequality Measure}
\author{Luke A. Prendergast and Robert G. Staudte}
\date{10 March, 2016}
\maketitle
\abstract{Ratios of quantiles are often computed for income distributions as rough measures of inequality, and inference
for such ratios have recently become available.  The special case when the quantiles are symmetrically chosen; that is, when the $p/2$ quantile is divided by the $(1-p/2)$, is of special interest because the graph of such ratios, plotted as a function of $p$ over the unit interval, yields an informative inequality curve. The area above the curve and less than the horizontal line at one is an easily interpretable coefficient of inequality. The advantages of these concepts over the
traditional Lorenz curve and Gini coefficient are numerous:  they are defined for all positive income distributions, they can be robustly estimated and distribution-free confidence intervals for the inequality coefficient are easily found. Moreover the inequality curves satisfy a  median-based transference principle and are convex for many commonly assumed income distributions..}

\bigskip


\vspace{2cm}

 \section{Introduction}\label{sec:intro}

\subsection{Background}
The widespread plotting of Lorenz curves and reporting of the associated Gini coefficients for income data
since their introduction \citep{lorenz-1905,gini-1914} over a century ago guarantees their historical importance. These original works also stimulated hundreds of theoretical papers. However, despite substantial progress in inferential
methodology, \citep[see][and references therein]{beach-1983,cowell-2003,davidson-2008}, there are inherent defects in the original concepts which preclude distribution-free methods. The first is the explicit requirement that the population mean exists (together with the implicit requirement that its variance exists in order to carry out inference). The second defect in these traditional concepts down-weight smaller incomes, thus giving too much emphasis to
the middle incomes.   The first defect can be overcome by utilizing quantile versions of the Lorenz curve which have recently been studied by \cite{PR&ST15}, and both defects can be overcome by employing the simple ratios of symmetric quantiles, which we now investigate.
    There are other {\em desiderata} that many economists might require of a measure of inequality, such as
mean-income tranference principles, \cite{cowell-2002}, and decomposability, \cite{bourg-1979}.  The first property
is not satisfied by the quantile measures, but they do preserve a parallel transference principle that is median preserving.
    Finally, it is important that inequality measures and their estimates be applicable to a wide range of income distributions.  Recent emphasis  has been on combinations of lognormal for the lower portion of incomes, with Pareto tails
for the upper; for recent examples and discussion see \cite{clem-2005}, \cite{ghosh-2011} and \cite{bee-2014}.  Another advantage
of the inequality measure described herein
is that it requires no parametric model assumptions.

\subsection{Definitions and basic properties}
Let $F$ satisfy $F(0-)=0$ and the {\em $p$th quantile} $x_p=Q(p)=F^{-1}(p)=\inf \{x:\ F(x)\geq p\}$, $0<p<1$.
Define the {\em symmetric ratio of quantiles} for $0 <p < 1$ by
$ R(p)=x_{p/2}/x_{1-p/2}$.  Clearly for each $p$, $R(p)$ gives the ratio of the typical (median) income of the lowest
proportion $p$ of incomes to the typical (median) income of the largest proportion $p$. Extend $R$ to [0,1] by defining $R(0)=0$ and $R(1)=1$.
The graph $\{(p,R(p)\}$ of $R$ has the following properties, as the reader can readily verify:
\begin{enumerate}
  \item $0\leq R(p)\leq 1$
  \item $R(p)$ is monotone increasing from $R(0)=0$ to $R(1)=1.$
  \item $R(p)=1$ for all $0<p<1$ if and only if all incomes are equal.
  \item $R(p)$ is scale invariant.
  \item After any median preserving transformation of funds from
         the upper half of incomes to the lower half of incomes, $R(p)$ can only increase.
\end{enumerate}
Define the {\em ratio coefficient of inequality} by $I=I(F)= 1- \int _0^1 R(p)dp$. Then $0\leq I\leq 1$ with $I(F)=0$ when
all incomes are equal. Letting $m=x_{0.5}$ and making the change of variable $x=F^{-1}(p/2)$ one obtains the
 following result:
\begin{equation}\label{coeff}
 1-I(F)=\int _0^1R(p)\;dp=\int _0^m\frac {2x\,dF(x)}{F^{-1}(1-F(x))}=\e \left [\frac{X}{Y} \right ]~,
\end{equation}
where $X\sim F(\cdot |X\leq m)$ and $F(X)+F(Y)=1$ defines $Y$.   If one selects an income at random from
those below the median and divides it by its symmetric quantile, on average one obtains $1-I(F).$  Therefore,  $I(F)$  has the simple interpretation as the {\em average relative distance $(Y-X)/Y$ of $X$ from its symmetric quantile $Y$}.

The useful properties for $I$ lead us to explore the measure as an alternative to the Gini Index which is defined as
\begin{equation}
G = 1 - \frac{1}{E(X)}\int^\infty_0[1- F(x)]^2dx.\label{G}
\end{equation}
Like $I$, $G\in [0,1]$ where $G=1$ equates to maximum inequality and $G=0$ for the situation of equal incomes for all.

\subsection{Summary of results}
We begin in Section~\ref{sec:examples} with  examples of  inequality curves and coefficients of inequality for some
common income distributions and compare values of $I$ with $G$.
Then in Section~\ref{sec:inference} we introduce empirical versions of these
concepts and investigate their inferential properties, including robustness to outliers.
 In particular large sample distribution-free confidence intervals
are obtained and their properties compared with those for the Gini Index. Applications to income data are in Section~\ref{sec:applications}.
A summary and discussion of further possible work is contained in Section~\ref{sec:summary}.  Although
convexity of the inequality curves is not considered by us to be a requisite for measuring inequality, it is an inherent property of Lorenz curves, and possessed for the ratios of symmetric quantiles for many distributions, so  conditions for convexity are given in Appendix~\ref{sec:convex}.

\begin{figure}[t!]
\begin{center}
\includegraphics[scale=.7]{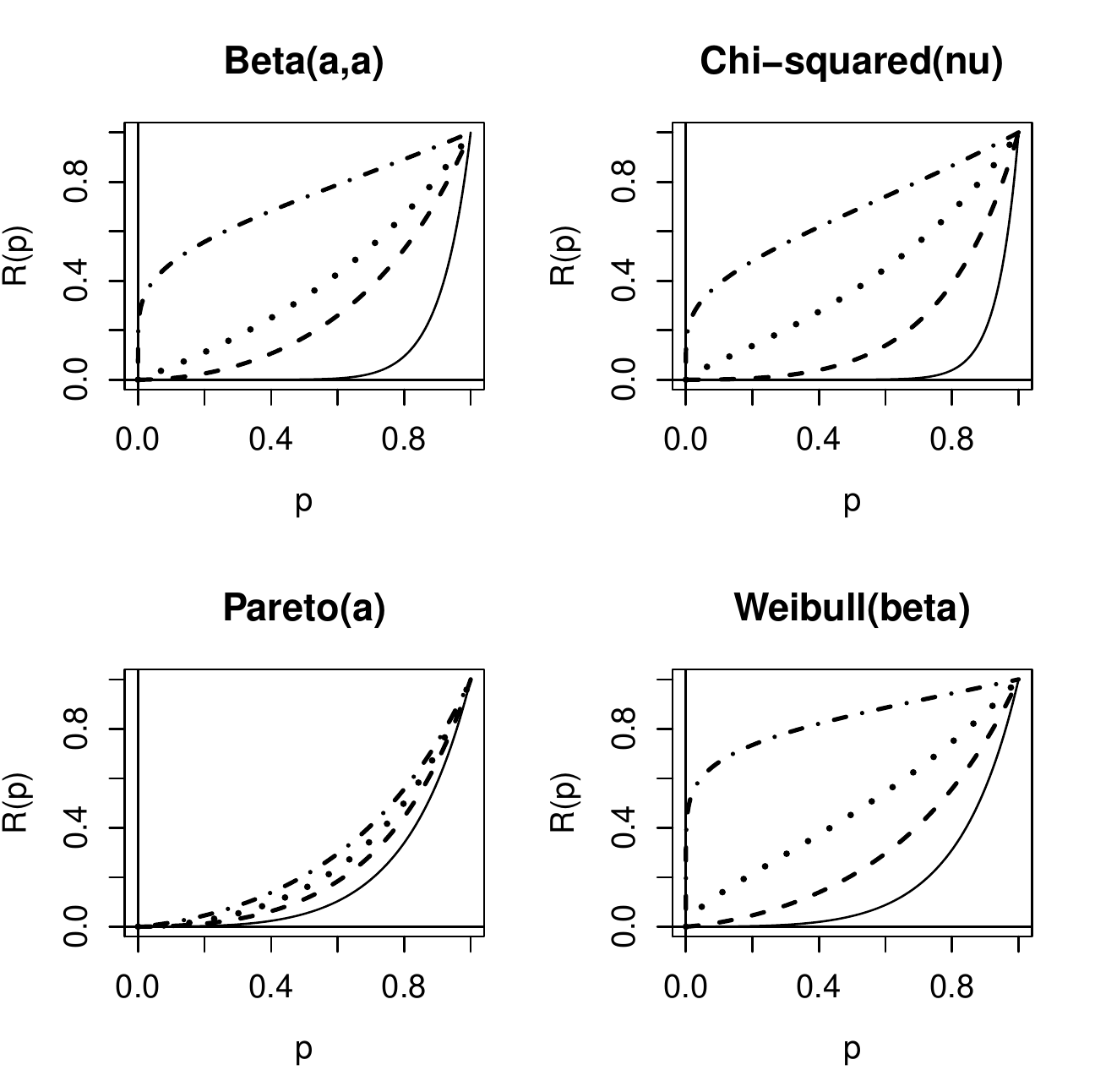}
\caption{\footnotesize \em Plots of $\{p,R(p)\}$ for some common distributions.  The top left graphs arise from the
symmetric Beta($a,a$) models with  parameters $a=0.1, 0.5, 1$ and 2, respectively, in solid, dashed, dotted and dash-dotted lines.   The top right graphs arise from Chi-squared$(\nu )$  models with  $\nu = 1/4, 1, 4$ and 25, respectively, in solid, dashed, dotted and dash-dotted lines.  The bottom left is for Type II Pareto$(a)$ with $a=1/2, 1, 2$ and 100, respectively; and the bottom right for Weibull$(\beta )$, with parameters $\beta =  1/2, 1, 2$ and 10.
\label{fig1}}
\end{center}
\end{figure}

\section{Examples}\label{sec:examples}

In this section we consider examples of the inequality curves $\{(p,R(p)\}$ and the associated
quantile inequality index $I$ for several well-known distributions before making some comparisons between $I$ and the Gini Index $G$. For background material on all the standard probability models in this paper, see
\cite{J-K-B-1994,J-K-B-1995}.

\subsection{Examples of the graphs of $R(p)$}

 In Figure~\ref{fig1} are shown the graphs $\{(p,R(p)\}$ of ratios of symmetric quantiles for some common probability models.
The area between each graph and the horizontal
 line at 1 is equal to $I(F)$, as described in (\ref{coeff}).

For the symmetric Beta$(a,a)$ family the densities are U-shaped for small $a$ with limiting case as $a\to 0$
of half the mass moving to each of the two points 0 and 1; and for large $a$ the densities
are unimodal with the limiting case as $a\to \infty $ having all the
incomes at one point 1/2.  Therefore the values of $I_a$ decrease from 1 to 0 as $a$
increases from 0 to $+\infty $. Note that  Beta(1,1) is the uniform distribution on [0,1].

 The Chi-squared($\nu$) densities become increasingly skewed as $\nu \to 0$, and
 $I_\nu $ decreases from 1 to 0 as $\nu $ increases from 0 to $+\infty $.
  Similarly, for the Weibull($\beta$) family, $I_\beta $ decreases from 1 to 0 as $\beta $
increases from 0 to $+\infty $ The  Weibull(1) is the exponential distribution, which has $R(p)= \ln (2-p)/\ln (p).$

For the Type II Pareto$(a)$ family, the graph of $R_a$ approaches that of the exponential  as $a \to +\infty $.
 Further, the range of $I_a$ is much more restricted than for the other families shown above,
 decreasing from 1 to 0.702 as $a$ increases from 0 to $+\infty .$ This family of income distributions only represents relatively high inequality.

The quantile function of the lognormal model is given by $Q(p)=e^{z_p}$ where $z_p=\Phi ^{-1}(p)$ is the quantile function of the standard normal.  Hence $R(p)= \exp \{z_{p/2}-z_{1-p/2}\}=\exp (2z_{p/2}).$
Its graph (not shown) is not unlike $\{(p,p^2)\}$ and has coefficient $I =0.6638.$  Moreover, although the graph appears to be convex, it is not convex for $p<0.045,$ as the reader can readily verify.  The convexity of $R$ cannot always be easily determined by inspection of a plot, so a more formal approach to convexity is taken in Appendix~\ref{sec:convex}.

\begin{figure}[h!t]
\begin{center}
\includegraphics[scale=.7]{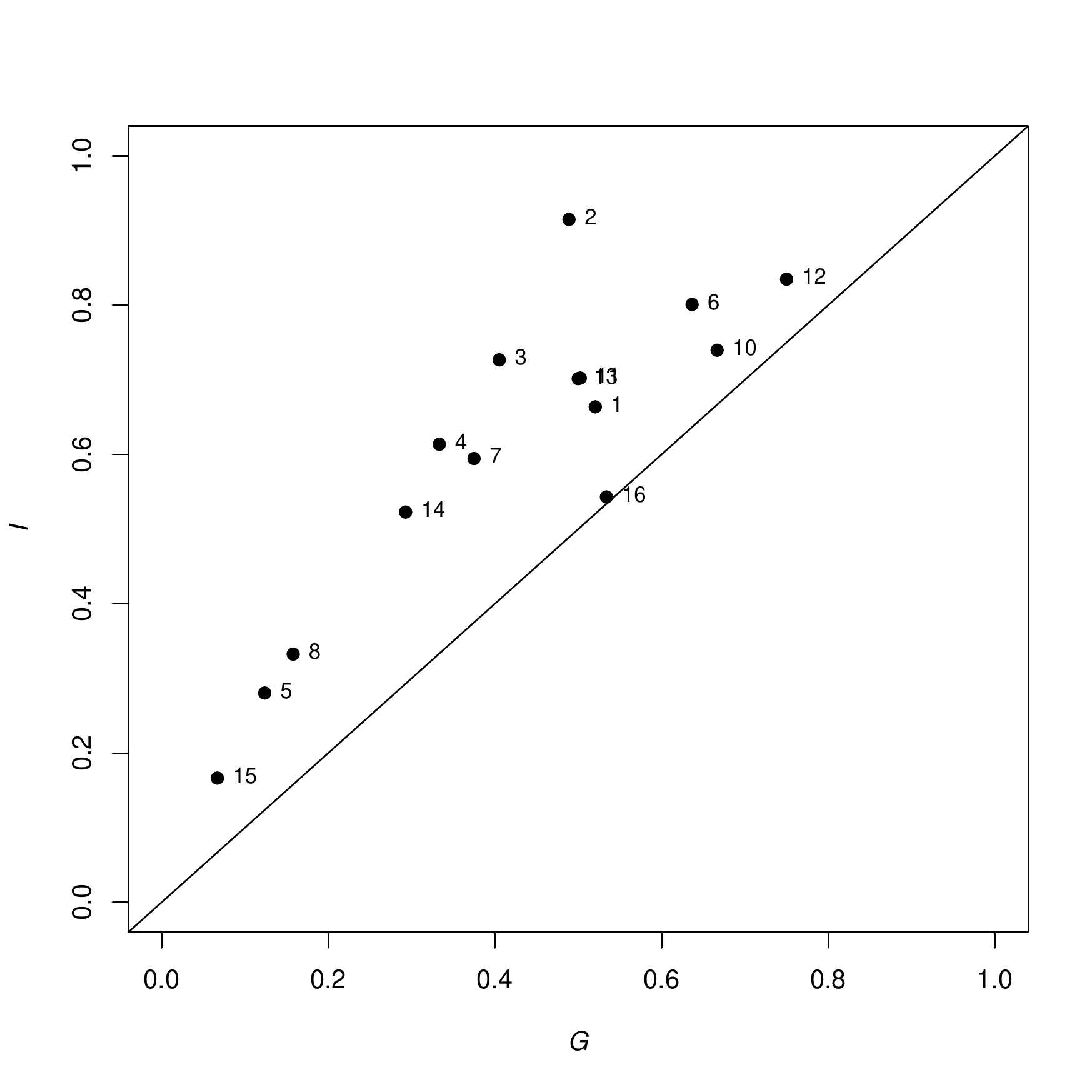}
\caption{\footnotesize \em A plot of the quantile inequality index $I$ against the Gini Index $G$ for the distributions considered in Table \ref{table1}. The numbers labeling the points identify the distributions listed in the table. \label{figgvi}}
\end{center}
\end{figure}

\begin{table}[h!t]
\begin{footnotesize}
\begin{center}
\caption{\label{table1} {\footnotesize \em Values of  $G$, $I$  computed numerically using adaptive quadrature over 1000 intervals. The Pareto distributions are Type II.  The parameter values for the compound lognormal-Frechet distribution were set approximately equal to those used in Illustration 2 of \cite{NA13}. The columns labelled $I^{(J)}$ are defined and interpreted in Section~\ref{sec:approxI}. } }
\vspace{.2cm}
\begin{tabular}{lllccccc}
\toprule
\#& $\qquad F$     & $G$     & $I$  & $I^{(50)}$ & $I^{(100)}$ & $I^{(500)}$   \\[.2cm]
\hline
1&Lognormal        &0.5205     & 0.6638  & 0.6638 & 0.6638& 0.6638\\
2&Beta(0.1,0.1)    &0.4889     & 0.9149  & 0.9150 & 0.9149& 0.9149\\
3&Beta(0.5,0.5)    &0.4053     & 0.7268  & 0.7268 & 0.7268& 0.7268\\
4&Beta(1,1)        &0.3333     & 0.6137  & 0.6137 & 0.6137& 0.6137\\
5&Beta(10,10)      &0.1238     & 0.2804  & 0.2802 & 0.2803& 0.2804\\
6&$\chi ^2_1$      &0.6366     & 0.8010  & 0.8011 & 0.8011& 0.8010\\
7&$\chi ^2_4$      &0.3750     & 0.5946  & 0.5946 & 0.5946& 0.5946\\
8&$\chi ^2_{25}$   &0.1580     & 0.3326  & 0.3324 & 0.3325& 0.3326\\
9&Pareto(1)        &0.9973*    & 0.7726  & 0.7727 & 0.7726& 0.7726\\
10&Pareto(2)       &0.6667     & 0.7397  & 0.7397 & 0.7397& 0.7397\\
11& Pareto(100)    &0.5025     & 0.7024  & 0.7024 & 0.7024& 0.7024\\
12& Weibull(0.5)   &0.7500     & 0.8348  & 0.8349 & 0.8349& 0.8348\\
13& Weibull(1)     &0.5000     & 0.7016  & 0.7016 & 0.7016& 0.7016\\
14& Weibull(2)     &0.2929     & 0.5229  & 0.5228 & 0.5229& 0.5229\\
15& Weibull(10)    &0.0670     & 0.1665  & 0.1662 & 0.1664& 0.1665\\
16& Lognormal-Frechet* & 0.5338& 0.5431  & 0.5431 & 0.5431& 0.5431\\ [.2cm]
\bottomrule
\end{tabular}
\end{center}
\end{footnotesize}
\end{table}

\subsection{Comparing values of $I$ with $G$}

In Figure \ref{figgvi} we plot examples of $I$ versus $G$ for the distributions listed in Table \ref{table1}.  \lq Lognormal-Frechet\rq\  refers to a composite lognormal, Frechet distribution, which is popular in income modeling, see \cite{NA13} for example.  The value for $G$ for the Pareto(1) distribution is marked with an asterisk since it is undefined (since the mean of the Pareto Type II is only defined for shape parameters greater than one).  However, it was computed numerically and we leave it for the purpose of comparisons later.  There are
clear differences between $G$ and $I$, but they are positively correlated, and  can be interpreted similarly in terms of
 rankings.
 A notable point of difference occurs for Model~2, the highly U-shaped
Beta(0.1,0.1) distribution, where $G\approx 0.49$ while $I\approx 0.91;$ this is a situation where the income population is
essentially composed of two equal size groups, having quite different values, and we think that $I$ better captures this disparity.

 A distinct advantage of $I$ over $G$ is that its values are more spread out, allowing for easier comparisons based on sample estimates, which we now demonstrate.

\section{Inference}\label{sec:inference}

In this Section we restrict attention to $F\in \F '$, where
\begin{equation*}
   \F '=\{F\in \F :\ f=F' \text { exists and is strictly positive.} \}
\end{equation*}
For such $F$ we can define the {\em quantile density} $q(p)=Q'(p)=1/f(x_p)$ of \cite{ par-1979}, which is
 also the {\em sparsity index} of \cite{tukey-1965}.

\subsection{Approximating $I(F)$}\label{sec:approxI}

 We next define a simple method for approximating $I$ that will prove useful in
the inference Section~\ref{sec:est}.  Given an integer $J$, define a grid $\{p_j\}$ on (0,1) by $p_j=(j-1/2)/J$, for $j=1,2,\dots ,J.$
 Then evaluate the ratio  $R(p_j)$ for $p_j$ in the grid and find $I^{(J)}\equiv  J^{-1}\,\sum _j\{1 -R(p_j)\}$.
One can make $I^{(J)}$ as close to $I$ as desired by choosing $J $ sufficiently large.

Table~\ref{table1} lists values of the Gini Index $G$, the quantile inequality index $I$ and the approximations to $I$ denoted $I^{(J)}$ for several choices of $J$. As can be seen, $I^{(J)}$ converges quickly with no differences to three decimal places reported between $I^{(100)}$ and $I^{(500)}.$

 Another example where an exact result can be obtained is the Type II Pareto distribution with shape parameter $a=1$ so that $Q(p)=1/(1-p)-1$.  We then have $1-\int^1_0R(p)dp=4\ln(2)-2=0.7726$ which again agrees, to four decimal places, with the results for $I^{(J)}$ reported in Table~\ref{table1}.

\subsection{Estimation}\label{sec:est}

Given a sample $X_1,\dots ,X_n$ from $F$ with order statistics $X_{(1)},\dots ,X_{(n)}$ and empirical cdf $F_n$
one can estimate $Q(p) $ by $Q_n(p)=F_n^{-1}(p)=X_{([np]+1)}$, but this is discontinuous in $p$, so we utilise
instead  the \cite{hynd-1996} estimator $\hat x_p$, which is a linear combination of two adjacent order statistics,
This estimator is Type~8 of quantile estimators on the software package R, \cite{R}.

For each $0<p<1$ define $\hat R(p)=\hat x_{p/2}/\hat x_{1-p/2}$. The asymptotic normality of arbitrary ratios of
sample quantiles $\hat x_p/\hat x_q$ is derived in \cite{ps-2015b}, so we only state that it can be shown
that $\sqrt n\,\{\hat R(p)-R(p)\}$ converges in distribution to a normal distribution with mean 0 and
variance $\sigma ^2_p$ given by:

\begin{equation}\label{ASVR}
  \sigma ^2_p = a_0 + a_1\; R(p) + a_2\; R^2(p)
\end{equation}
where $a_0=(p/2)\,(1-p/2)\,q^2(p/2)/x_{1-p/2}^2$, $a_1=-2(p/2)^2\,q(p/2)\,q(1-p/2)/x_{1-p/2}^2$ and $a_2=(p/2)\,(1-p/2)\,q^2(1-p/2)/x_{1-p/2}^2$.  This formula enables one to find large-sample 100$(1-\alpha )$\% confidence intervals for $R(p)$ of the form $\hat R(p)\pm z_{1-\alpha/2}\hat\sigma _p/\sqrt{n}$, where $\hat \sigma _p$ requires estimates of the quantile density $q$ at $p/2$ and $1-p/2$; specific estimates are derived in \cite{ps-2016} and utilized in \cite{ps-2015b}. Here we want confidence intervals for $I=1-\int R(p)\,dp$, which we estimate by  $\hat I=1-\int \hat R(p)\,dp.$
Closed form expressions for $I$ and $\hat I$ are not usually available, so we obtained numerical approximations
to them as follows.

  As in Section \ref{sec:approxI} where we approximated $I$ by $I^{(J)}$,  we estimate $I^{(J)}$, and hence $I$, as follows. Let $p_j=(j-1/2)/J$, for $j=1,2,\dots ,J.$ and for each $j$ let $\hat R_i(p_j)$ be the estimated ordinate of the ratio inequality curve  at $p_j$.  Then $\widehat I^{(J)}$ is defined by
 \begin{equation}\label{Ihat}
 \widehat I^{(J)}\equiv (1/J)\, \sum _j\{1-\hat R(p_j)~.
 \end{equation}
 \cite{beach-1983} find the limiting {\em joint} normal distribution of estimates of a finite number of Lorenz curve ordinates, based on a finite number of sample quantiles, assuming $F\in \F'$ has a finite mean. In the same way, for $F\in \F '$, the  limiting joint normal distribution of the estimated ordinates $\hat R(p_j)$, $j=1,\dots ,J$ can be established. An analytic expression for the covariance matrix is not required by us, only  asymptotic normality of $\hat I$, which being an average of the $\hat R(p_j)$s, is immediate. An approximate variance of $\widehat I^{(J)}$, $\text{Var}[\widehat I^{(J)}]$, is given in \eqref{VarI} in the Appendix, and an asymptotic $(1-\alpha)\times 100$ confidence interval for  $I^{(J)}$ is

 \begin{equation}\label{CI}
[L,U]=\widehat I^{(J)} \pm z_{1-\alpha/2}\sqrt{\text{Var}[\widehat I^{(J)}]}~,
 \end{equation}
 where $z_{1-\alpha/2}=\Phi^{1}(1-\alpha/2)$ is the $1-\alpha/2$ quantile from the standard normal distribution.  Similarly, if we have two estimated $I$'s, $\widehat{I}_1^{(J)}$ and $\widehat{I}_2^{(J)}$, arising from two independent samples, then an interval estimate for the difference between the two is

 \begin{equation}\label{CI2}
[L,U]_\text{diff}= \widehat I_1^{(J)} - \widehat I_2^{(J)} \pm z_{1-\alpha/2}\sqrt{\text{Var}[\widehat I_1^{(J)}]+\text{Var}[\widehat I_2^{(J)}]}~,
 \end{equation}
 where, for simplicity,  the same $J$ is used for estimates of the inequality index.

\subsection{Interval coverage and width}\label{sec:ci}

In this section we assess the coverage probability and expected interval width for the interval estimators of $I$ given in \eqref{CI}.  We also provide some comparisons with estimators of $G$ where the $(1-\alpha/2)\times 100\%$ confidence interval for $G$ is computed as $\widehat{G}\pm z_{1-\alpha/2}\sqrt{\text{Var}(\widehat{G})}$, and where $\text{Var}(\widehat{G})$ is given in \cite{davidson-2008}.

In Table~\ref{table2} are listed simulated coverage probabilities and average widths for the intervals (\ref{CI}) for several choices of sample sizes and the same wide range of distributions.  Of particular merit is the fact that the coverage probability is, in most cases, slightly above the nominal 0.95.  When it is below 0.95, the coverage is still at least very good with the smallest coverage found to be 0.93 for the Beta(0.5,0.5) distribution when $n=100$.  Overall, the simulations suggest reliable coverage, even for $n=100$, and  narrow expected interval widths relative to $I$.  The distributions  differ enough to suggest that the interval estimator will be reliable in practice.  Similar results were found for other choices of $J$.

\begin{table}[ht]
\caption{\label{table2} {\footnotesize \em Empirical coverage probabilities and average widths of interval estimates (\ref{CI}) of $I$  at nominal level 95\%, each based on 4000 replications.  The grid points number $J=100$.} }
\vspace{0.2cm}
\centering
\begin{footnotesize}
\begin{tabular}{llcccccc}
\toprule
\#& Distribution  & $I$  & $n=100$      & $n=200$      & $n=500$       & $n=1000$      & $n=5000$     \\  \hline
1& Lognormal     &0.664 &0.973 (0.124) &0.965 (0.084) &0.962 (0.051)  &0.961 (0.036)  &0.955 (0.016) \\
2& Beta(0.1,0.1) &0.915 &0.921 (0.165) &0.947 (0.117) &0.963 (0.074)  &0.962 (0.052)  &0.958 (0.023) \\
3& Beta(0.5,0.5) &0.727 &0.930 (0.138) &0.934 (0.099) &0.940 (0.063)  &0.947 (0.045)  &0.944 (0.021) \\
4& Beta(1,1)     &0.614 &0.944 (0.135) &0.944 (0.095) &0.954 (0.060)  &0.950 (0.043)  &0.950 (0.019) \\
5& Beta(10,10)   &0.280 &0.963 (0.082) &0.967 (0.058) &0.969 (0.036)  &0.966 (0.025)  &0.959 (0.011) \\
6& $\chi ^2_1$   &0.801 &0.976 (0.120) &0.967 (0.081) &0.958 (0.050)  &0.952 (0.035)  &0.954 (0.015) \\
7& $\chi ^2_4$   &0.595 &0.959 (0.120) &0.961 (0.083) &0.957 (0.052)  &0.953 (0.036)  &0.954 (0.016) \\
8& $\chi ^2_{25}$&0.333 &0.960 (0.088) &0.967 (0.062) &0.969 (0.039)  &0.961 (0.027)  &0.960 (0.012) \\
9& Pareto(1)     &0.773 &0.984 (0.134) &0.979 (0.086) &0.967 (0.051)  &0.966 (0.035)  &0.959 (0.015) \\
10&Pareto(2)     &0.740 &0.977 (0.124) &0.970 (0.083) &0.962 (0.051)  &0.959 (0.035)  &0.956 (0.015) \\
11&Pareto(100)   &0.702 &0.964 (0.121) &0.956 (0.084) &0.954 (0.052)  &0.954 (0.036)  &0.951 (0.016) \\
12&Weibull(0.5)  &0.835 &0.990 (0.124) &0.980 (0.081) &0.973 (0.048)  &0.965 (0.033)  &0.956 (0.014) \\
13&Weibull(1)    &0.702 &0.962 (0.121) &0.958 (0.084) &0.954 (0.052)  &0.952 (0.036)  &0.955 (0.016) \\
14&Weibull(2)    &0.523 &0.960 (0.121) &0.963 (0.084) &0.961 (0.053)  &0.958 (0.037)  &0.953 (0.016) \\
15&Weibull(10)   &0.166 &0.965 (0.063) &0.972 (0.044) &0.979 (0.028)  &0.975 (0.019)  &0.962 (0.008) \\
16&LN-Frechet*   &0.543& 0.976 (0.167) & 0.986 (0.108) & 0.972 (0.062) & 0.960 (0.043) & 0.948 (0.018)\\ \bottomrule
\end{tabular}
\end{footnotesize}
\end{table}

In Table~\ref{table3} we repeat the simulations from Table~\ref{table2} but this time for interval estimation of $G$.  The extremely poor coverages for the Pareto(1) are expected since, as noted previously, $G$ is not defined.  While most of the coverages are reasonably close to the nominal level of 0.95, unlike the interval estimators for $I$ they are not consistently so.

\begin{table}[hb]
\caption{\label{table3} {\small \em Simulated coverage probabilities for the estimation of $G$ and mean interval widths for
nominal 95\% confidence interval estimates.} }
\vspace{0.2cm}
\centering
\begin{footnotesize}
\begin{tabular}{llcccccc}
\toprule
\#& Distribution  & $G$  & $n=100$      & $n=200$      & $n=500$       & $n=1000$      & $n=5000$     \\  \hline
1& Lognormal     &0.521& 0.869 (0.136) &0.893 (0.104) &0.919 (0.070)  &0.935 (0.051)  &0.951 (0.023)\\
2& Beta(0.1,0.1) &0.489& 0.951 (0.178) &0.949 (0.125) &0.949 (0.079)  &0.948 (0.056)  &0.952 (0.025)\\
3& Beta(0.5,0.5) &0.405& 0.951 (0.125) &0.947 (0.088) &0.952 (0.055)  &0.951 (0.039)  &0.951 (0.017)\\
4& Beta(1,1)     &0.333& 0.946 (0.097) &0.948 (0.068) &0.952 (0.043)  &0.948 (0.030)  &0.952 (0.013)\\
5& Beta(10,10)   &0.124& 0.930 (0.034) &0.945 (0.024) &0.948 (0.015)  &0.949 (0.011)  &0.952 (0.005)\\
6& $\chi ^2_1$   &0.637& 0.932 (0.120) &0.936 (0.086) &0.950 (0.055)  &0.947 (0.039)  &0.948 (0.017)\\
7& $\chi ^2_4$   &0.375& 0.934 (0.092) &0.943 (0.065) &0.945 (0.042)  &0.948 (0.030)  &0.951 (0.013)\\
8& $\chi ^2_{25}$&0.158& 0.935 (0.043) &0.942 (0.030) &0.948 (0.019)  &0.945 (0.014)  &0.948 (0.006)\\
9& Pareto(1)     &0.997*& 0.012 (0.139) &0.053 (0.130) &0.116 (0.116)  &0.115 (0.104)  &0.114 (0.079)\\
10&Pareto(2)     &0.667& 0.751 (0.162) &0.793 (0.136) &0.843 (0.104)  &0.864 (0.083)  &0.894 (0.045)\\
11&Pareto(100)   &0.502& 0.931 (0.111) &0.940 (0.080) &0.948 (0.051)  &0.950 (0.036)  &0.954 (0.016)\\
12&Weibull(0.5)  &0.750& 0.885 (0.123) &0.906 (0.093) &0.925 (0.063)  &0.938 (0.045)  &0.946 (0.021)\\
13&Weibull(1)    &0.500& 0.934 (0.110) &0.947 (0.079) &0.943 (0.050)  &0.949 (0.036)  &0.955 (0.016)\\
14&Weibull(2)    &0.293& 0.938 (0.077) &0.947 (0.054) &0.949 (0.034)  &0.947 (0.024)  &0.952 (0.011)\\
15&Weibull(10)   &0.067& 0.929 (0.021) &0.938 (0.015) &0.950 (0.010)  &0.948 (0.007)  &0.949 (0.003)\\
16&LN-Frechet*   &0.534& 0.617 (0.230) &0.674 (0.209) &0.720 (0.173)  &0.739 (0.147)  &0.775 (0.097)\\ \bottomrule
\end{tabular}
\end{footnotesize}
\end{table}

In Table \ref{table9} of the Appendix~\ref{sec:appbias} we make comparisons of the estimators of $G$ and $I$ by computing their empirical biases and standard errors.  Again, the large bias reported for estimation of $G$ in the case of the Pareto(1) distribution are not surprising since $G$ is not defined (and comparisons with a numerically computed, but incorrect, $G$ are made).  The difference in performance
of the estimators $\hat I$ and $\hat G$ are more dramatically revealed in the next Section.

\clearpage
\newpage

\subsection{Robustness properties}\label{sec:rob}

\cite{cowell-1996} show that the Lorenz curve ordinates and Gini index have unbounded influence functions, but the ratio of quantiles is well known to have a bounded influence, see \cite{ps-2015b}, for example.
This implies that the influence function of the quantile inequality index $I$ is also bounded, because it is an average of bounded influence functions.

\begin{figure}[h!]
\begin{center}
\includegraphics[scale=.75]{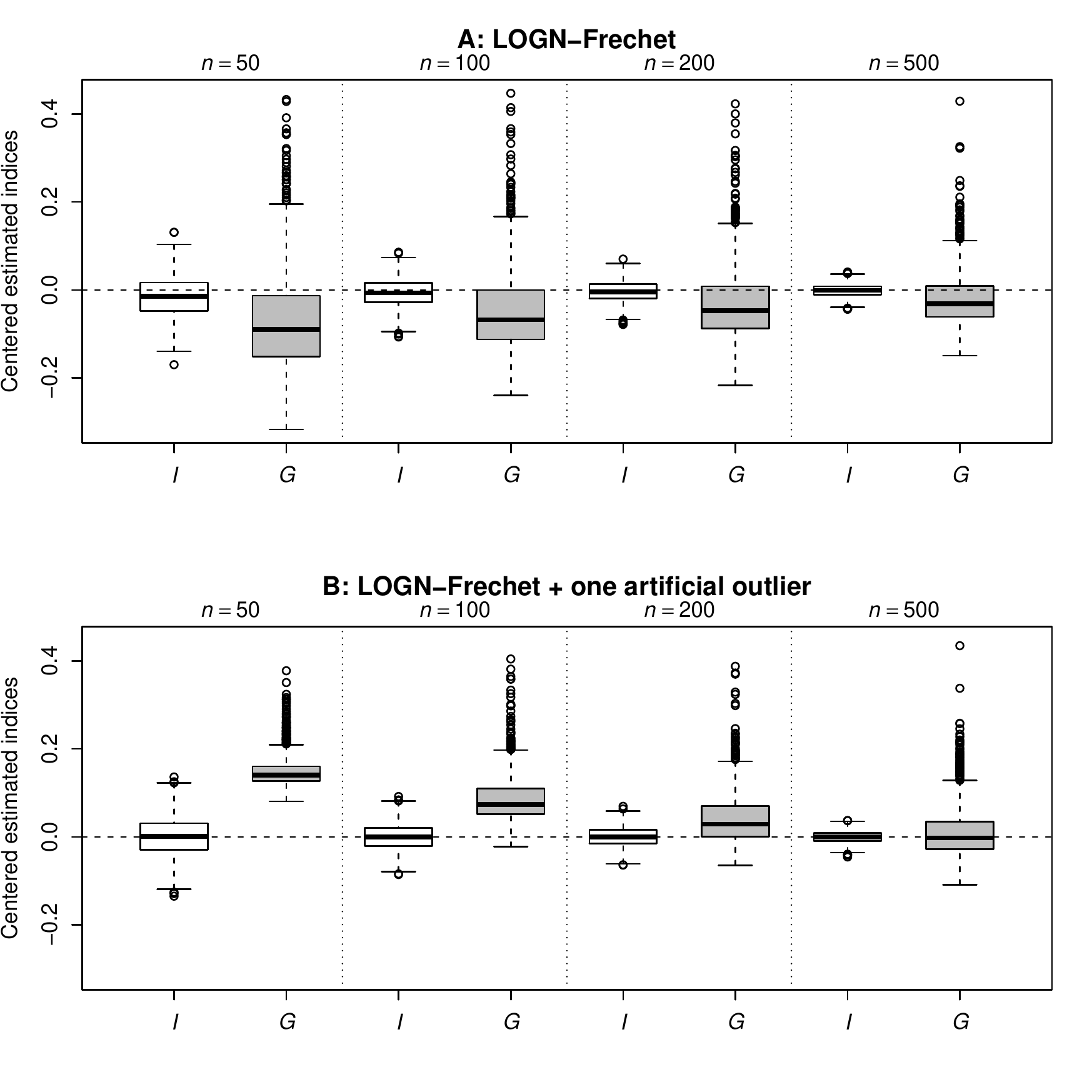}
\caption{\footnotesize \em Boxplots of 1000 centered (with respect to the true $I=0.5431$ and $G=0.5338$) simulated estimates of $I$ and $G$ from $n$ observations randomly generated from the composite lognormal-Frechet distribution with parameters $\exp(-1.72)$, $\exp(0.12)$, $\exp(-0.29)$ and $\exp(0.41)$ (Plot A).  In Plot B the simulation is repeated but where one observation is replaced with the 0.999 quantile as an outlier. \label{boxplots2}}
\end{center}
\end{figure}

In this section we provide simulations that show that $I$ is robust  and provides a better alternative to the Gini index $G$ when outliers are present.
In Figure \ref{boxplots2} we provide boxplots of 1000 simulated estimates of $I$ and $G$ from $n$ observations randomly generated from a composite lognormal-Frechet distribution \cite{NA13}.

Plot A includes estimates resulting from a straight random sampling from the distributions while in Plot B we replace one observation in the sampling with a large outlier (the 0.999 quantile). Even in Plot A where no artificial outlier was included, it can be seen the estimates of $G$ are biased.  Additionally, the variance in estimation of $G$ is large especially when compared to the much lower variability shown for the estimates of $I$.  Another benefit of $I$ is in the fact that the bias of the estimates is negligible.  Overall properties of the estimator of $I$ in this setting (small bias and small variability) suggest that it is an attractive choice as a measure of inequality.

\begin{figure}[h!t]
\begin{center}
\includegraphics[scale=.75]{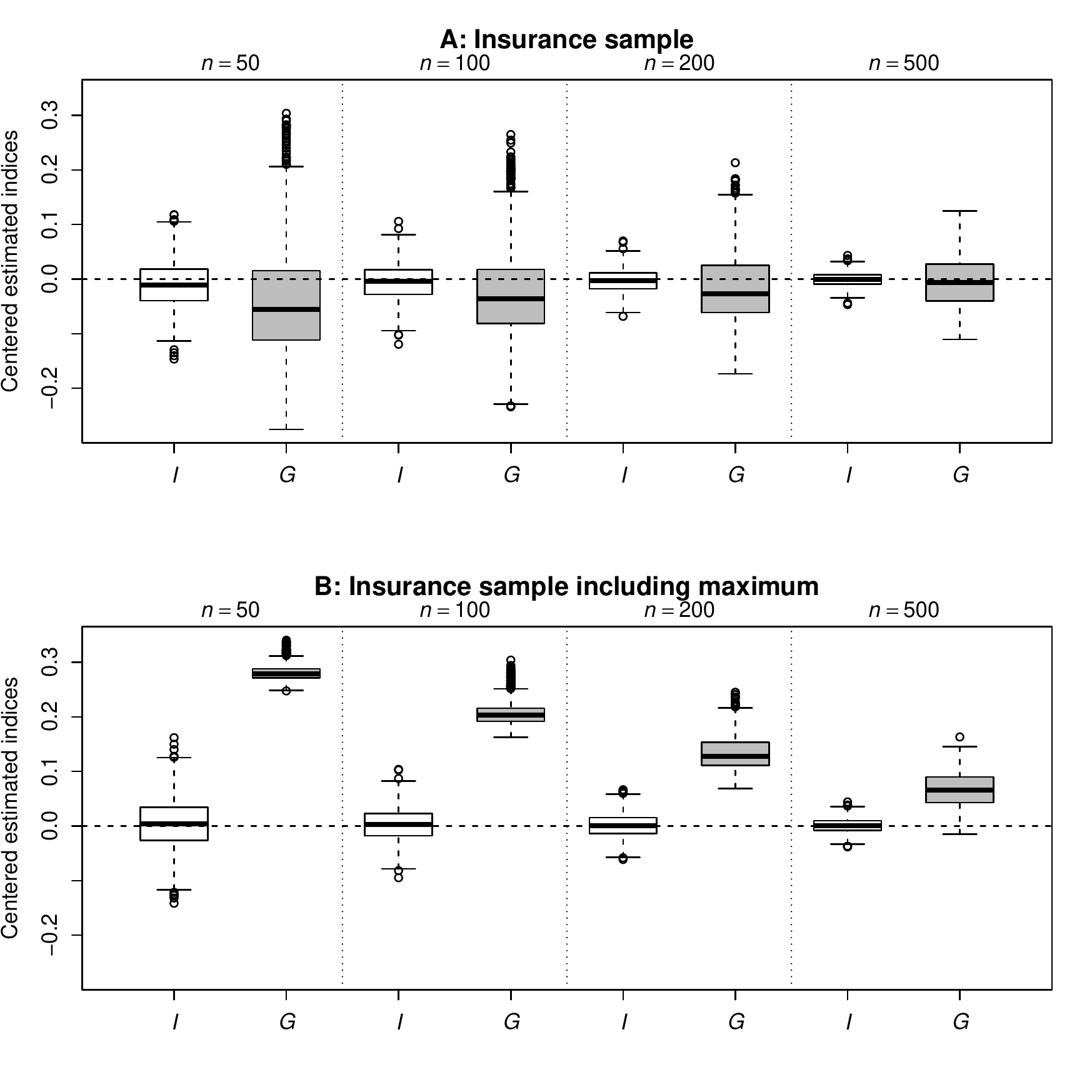}
\caption{\footnotesize \em Boxplots of 1000 centered (with respect to the true $I=0.5337$ and $G=0.5150$) simulated estimates of $I$ and $G$ from $n$ observations randomly selected from a real data set consisting of 2492 insurance claims in Denmark (Plot A).  In Plot B the simulation is repeated but where one observation is replaced with the maximum value from the complete data set.
\label{boxplots3}}
\end{center}
\end{figure}

We now repeat the simulation but sample from a data set that consists of 2492 Danish fire insurance claims; it
 is available from the R package \textit{SMPracticals} \citep{DA13} and is considered on Page 278 of \cite{DA13}.  Plot A depicts boxplots of 1000 simulated estimates for $n$ observations randomly selected from the complete data set.  The estimates are $\widehat{G}=0.5150$ and $\widehat{I}=0.5337$.  While the variability of $\hat G$ has decreased for smaller $n$ the bias is large when $n\leq 200$.
  Bias is not a problem for $\widehat{G}$ when $n=500$, but the variance is far greater than it is for the estimation of $I$.  Once again we see that the estimates of $I$ are excellent when compared to those for $G$, with smaller variability and bias.  In Plot B we repeat the simulation, but now  include in each sample the largest value in the original data set.   The estimator of $I$ is hardly affected by the outlier.  However, the estimator of $G$ is heavily biased which is notable even when $n=500$.

\clearpage
\newpage

\section{Applications}\label{sec:applications}

We next consider four applications to illustrate the versatility and effectiveness of $I$.

\subsection{Example 1: Danish insurance data}\label{ex1}

 We first estimate  $G$ and $I$ for the Danish insurance data introduced in the last section.  The estimate of $I$ is $\widehat{I}=0.5337$ which has small standard error of approximately 0.007 and subsequent 95\% confidence interval $(0.5204,\ 0.5470)$.  In contrast, while $\widehat{G}=0.5150$ is similar in magnitude, the standard error is much larger at 0.0230 resulting in much less certainty in the 95\% confidence interval estimator $(0.4699,\ 0.5601)$.

\subsection{Example 2: Earnings data}\label{ex2}

In this example hourly earnings data from 1992 (2962 paired (male-female) observations) and 1998 (2603 paired observations).  The data can be found as file a\texttt{CPSch3.csv} at
{\tt https://vincentarelbundock.github.io/Rdatasets/datasets.html}
and has been considered by \cite{ST&WA03}.

\begin{figure}[h!t]
\begin{center}
\includegraphics[scale=.5]{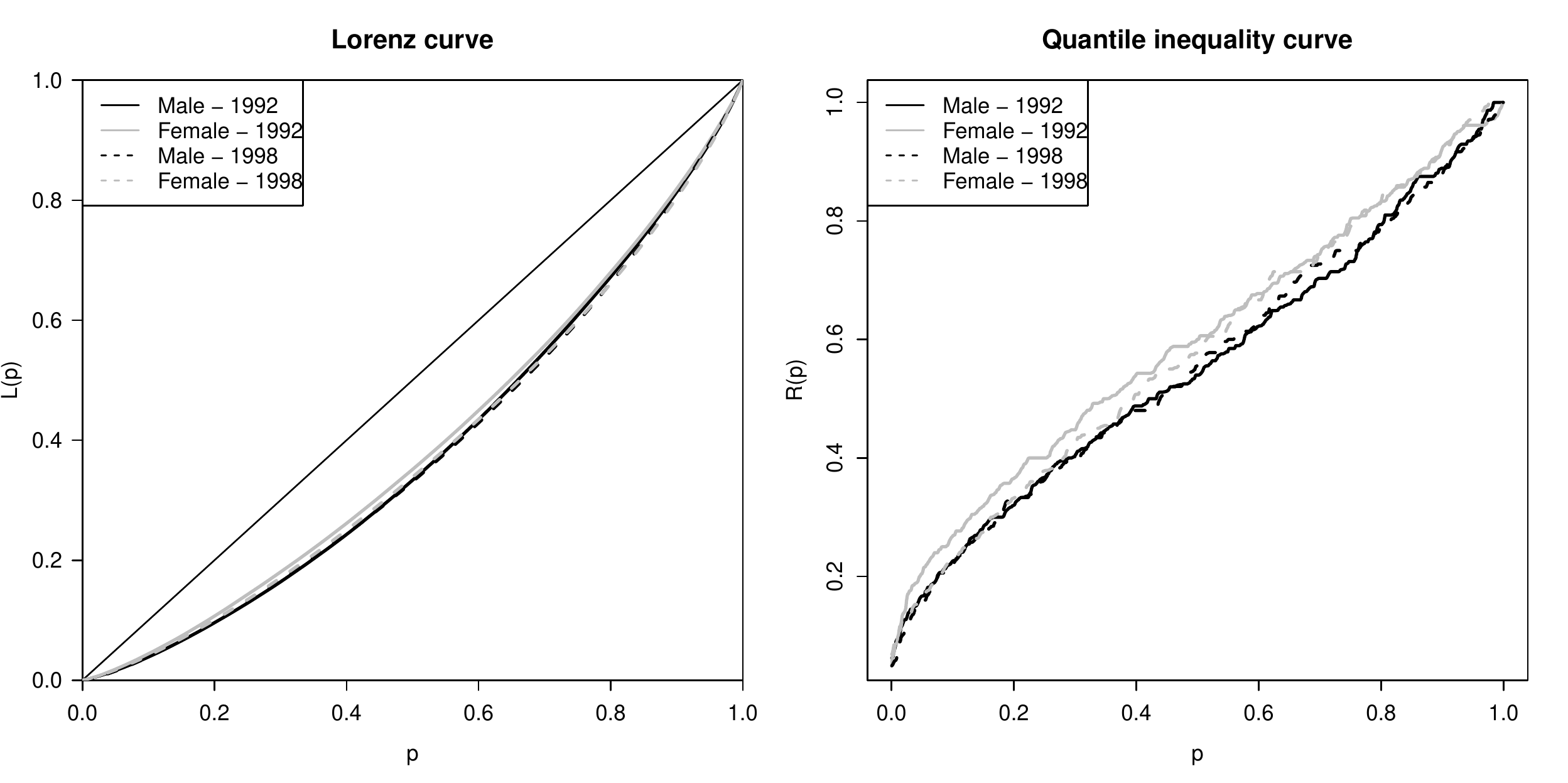}
\caption{\footnotesize \em Plots of the Lorenz curve and the quantile inequality curve for the earnings data for males and females in the years 1992 and 1998.
\label{fig:earnings}}
\end{center}
\end{figure}

Figure \ref{fig:earnings} depicts the Lorenz and quantile inequality curves for males and females in 1992 and 1998.  As can be seen, the Lorenz curves for each set of the data are very similar suggesting little difference in inequality.  The quantile inequality curve on the other hand tells a different story.  There is a difference between the curves for males and females in 1992 and in 1998 with stronger evidence for the former.  This suggest greater inequality among males when compared to females.

\begin{table}[ht]
\begin{center}
\caption{Point and interval estimates of $G$ and $I$ for earnings of males (M) and females (F) in 1992 and 1998 including differences between years (labeled 1998-1992) and between gender (labeled M-F).  CI-W refers to Wald-type intervals and CI-B to bootstrap intervals.\label{table4}}
 \vspace{0.3cm}
\begin{footnotesize}
\begin{tabular}{llccccccccc}
\toprule
      &            &\multicolumn{2}{c}{1992}                  &&\multicolumn{2}{c}{1998}        &&   \multicolumn{2}{c}{1998-1992}       \\
Gender&            &$G$         &$I$           &&$G$         &$I$         &&   $G$   &   $I$    \\
\cline{1-4} \cline{6-7} \cline{9-10}
M     &  Est.  &        0.235  &        0.446   &&        0.241  &        0.445  &&         0.006  &        -0.001    \\
      &  CI-W &(.227, .243) & (.433, .459) && (.232, .249)& (.431, .458)&& (-.006, .017)&  (-.02, .018)  \\
      &  CI-B &(.227, .243) & (.434, .459) &&  (.233, .25)& (.432, .457)&& (-.007, .019)& (-.017, .016)  \\
      \cline{1-4} \cline{6-7} \cline{9-10}
F     &  Est. &    0.214      &    0.404       &&    0.236      &     0.42      &&     0.022      &     0.017        \\
      &  CI-W &(.205, .222) & (.39, .417)  &&(.226, .245) &(.405, .435) && (.009, .035) &(-.003, .036)   \\
      &  CI-B &(.205, .222) & (.391, .417) && (.226, .246)& (.407, .433)&&  (.009, .034)& (-.001, .037)  \\
      \cline{1-4} \cline{6-7} \cline{9-10}
M-F   &  Est.&     0.022     &     0.042      &&     0.005     &     0.025     &&     -0.016     &     -0.018       \\
      &  CI-W&(.01, .033)  & (.024, .061) &&(-.007, .018)& (.005, .045)&& (-.034, .001)&  (-.045, .01)  \\
      &  CI-B&  (.01, .031)& (.025, .058) &&(-.008, .018)& (.006, .043)&&               -&               -  \\
\bottomrule
\end{tabular}
\end{footnotesize}
\end{center}
\end{table}

In Table~\ref{table4} we provide estimates of $G$ and $I$ for the earnings data for males (M) and females (F) in 1992 and 1998 including differences between years (labeled 1998-1992) and between gender (labeled M-F).  As a matter of comparison we give the Wald-type intervals such as those in \eqref{CI} and \eqref{CI2} as well as bootstrap intervals with 500 replicates.  The bootstrap intervals are taken to be the 2.5\% and 97.5\% quantiles of the 500 bootstrapped estimates of $G$ and $I$.  As can be seen the Wald-type and bootstrap intervals are almost identical.  The interval estimates for the difference between males and females in 1992 based on $G$ and $I$ both indicate a difference in 1992.  With respect to $G$, this was difficult to ascertain from the Lorenz curve alone.  The evidence is more compelling according to $I$.  In 1998, there was no difference between males and females according to $G$ with an estimated difference close to zero.  However, a difference is found for $I$ and although reasonably small, the interval estimate agrees with our notion that the quantile curves were different between the genders.

\subsection{ABS weekly income data}\label{ex3}

\begin{table}[b!]
\begin{footnotesize}
\begin{center}
\caption{Australian gross household weekly income (GWI) and (equivalized)
disposable weekly income (DWI) data for years 1995 and 2010 \cite{ABSincomedata}, Document 6523.0.\label{table5}}
 \vspace{0.3cm}
 \begin{tabular}{rrrrrrr}
  \multicolumn{3}{c}{Number of households ('000)}  && \multicolumn{3}{c}{Number of persons ('000)}\\
  GWI           & 1995 &   2010      &&  DWI  &   1995 &   2010 \\
\hline
    \$1-\$99    &  72.9 &    81.6     &&    \$1-\$49      &   132.0    &   102.0  \\
  \$100-\$199   &  75.5 &    62.3     &&    \$50-\$99     &    97.7    &    65.3  \\
  \$200-\$299   & 670.4 &   170.7     &&    \$100-\$149   &   170.2    &   101.7  \\
  \$300-\$399   & 371.5 &   645.0     &&    \$150-\$199   &   301.1    &   154.5  \\
  \$400-\$499   & 555.5 &   328.8     &&    \$200-\$249   &  1193.5    &   273.6  \\
  \$500-\$599   & 375.4 &   497.6     &&    \$250-\$299   &  1768.3    &   463.0  \\
  \$600-\$799   & 668.3 &   802.4     &&    \$300-\$349   &  1550.8    &  1150.7  \\
  \$800-\$999   & 589.0 &   637.2     &&    \$350-\$399   &  1509.0    &  1319.9  \\
  \$1000-\$1199 & 552.9 &   605.2     &&    \$400-\$449   &  1175.1    &  1091.9  \\
  \$1200-\$1399 & 509.6 &   556.0     &&    \$450-\$499   &  1246.7    &  1101.8  \\
  \$1400-\$1599 & 408.8 &   513.5     &&    \$500-\$599   &  2232.6    &  2283.5  \\
  \$1600-\$1799 & 325.8 &   469.3     &&    \$600-\$699   &  1948.3    &  2278.3  \\
  \$1800-\$1999 & 289.8 &   445.4     &&    \$700-\$799   &  1280.2    &  1868.4  \\
  \$2000-\$2499 & 525.7 &   869.5     &&    \$800-\$899   &   903.6    &  1745.2  \\
  \$2500-\$2999 & 211.2 &   566.3     &&    \$900-\$999   &   671.4    &  1492.5  \\
  \$3000-\$3999 & 162.9 &   634.1     &&    \$1000-\$1099 &   419.5    &  1196.0  \\
  \$4000-\$4999 &  46.6 &   230.0     &&    \$1100-\$1399 &   549.0    &  2359.6  \\
  \$5000 or more&  57.7 &   243.2     &&    \$1400-\$1699 &   165.9    &  1107.6  \\
                &       &             &&   \$1700-\$1999  &    49.4    &   573.3  \\
                &       &             &&   \$2000 or more &    73.1    &   771.7  \\
 \end{tabular}
 \end{center}
  \end{footnotesize}
   \end{table}

Table~\ref{table5} shows estimates of the gross weekly incomes of Australian households (GWI) and the disposable personal weekly incomes (DWI) for the years 1995 and 2010. We have omitted the categories
of \lq negative income\rq\  and \lq no income\rq,\ because their effect on our analysis is negligible compared to the right tail of large incomes.  Note in particular that for the 1995 GWI data the
last category \lq \$5000 or more\rq\ has lower bound $x_q=5000$, where
$q=1-57.7/(72.9+\dots +57.7)=0.991.$  We do not want to ignore this category.

Lacking the individual data, we can create an (admittedly {\em ad hoc})  population to take samples from.  We do this by generating uniformly distributed variables over each category, starting with
729 observations uniformly between 0 and 100, 755 observations uniformly between 100 and 200, and
so on.  For the last category we generate 577 random Pareto$(a,\lambda )$ observations follows:\quad first, for $a>0$, find the scale parameter $\lambda =x_q/\{(1-q)^{-1/a}-1\}=5000/
\{(1-0.991)^{-1/a}-1\}$; second, generate 577  $u_i$ from [0.991,1]; and third,
apply the quantile function to these values $Q_{a,\lambda }(u_i)=\lambda\{(1-u_i)^{-1/a}-1\}.$

\begin{figure}[t!]
\begin{center}
\includegraphics[scale=.8]{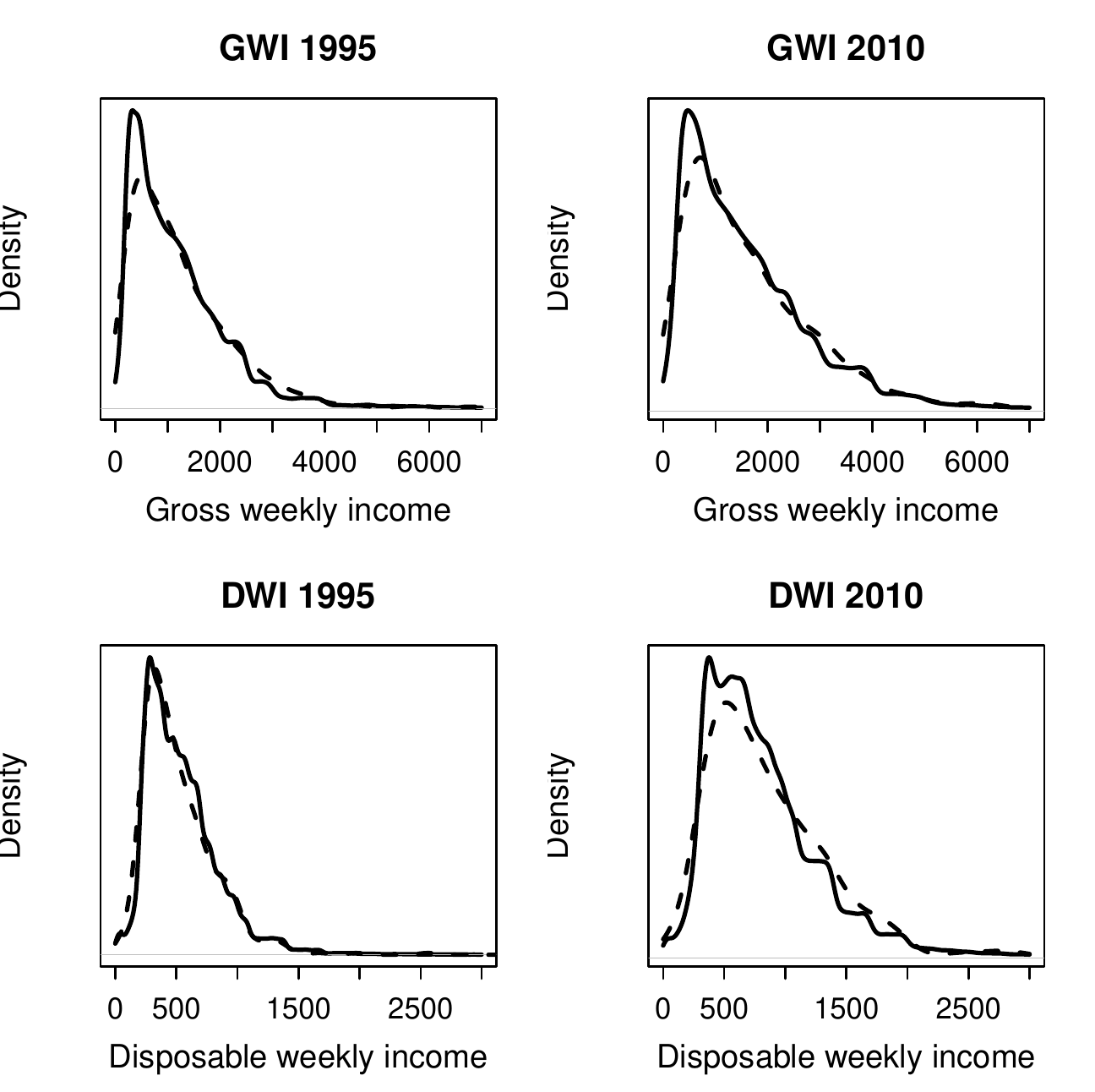}
\caption{\footnotesize \em Solid lines show kernel density plots of the populations constructed
from the categorical data in Table~\ref{table5} using Pareto $a=4$ tails. For details, see
Section~\ref{ex3}.
The dashed lines show density plots of samples of size 500 from each of the respective populations.
The rounded sample quartiles for the top left are 487, 985,  1653 amd the maximum was  6779.
For the top right they are 701, 1310,  2335 and the maximum 15490.  For the bottom left
 299,  444,   665 and  2560; while for the bottom right they are 496. 764, 1161 amd  9166.
\label{fig:absdata}}
\end{center}
\end{figure}

A kernel density estimate of this population data for $a=4$ is shown as a solid curve in the upper left plot of Figure~\ref{fig:absdata}.   The density plot has been truncated at 7000 but the maximum
value in this population is actually 27880. The quartiles are 476, 948 and 1602.
The other populations GWI-2010, DWI-1995 and DWI-2010, are similarly constructed and plotted for $a=4$.   Other populations were also constructed for $a=1, 2$ and 3 but are not shown; needless to say their outlying observations tend to be even larger.

\setlength{\tabcolsep}{4pt}
\begin{table}[h!]
\begin{center}
\begin{footnotesize}
\caption{{\bf Comparison of $\hat G$ and $\hat I$ for various tail shapes.}\em
Summary results $\hat G (\se[\hat G])$ and   $\hat I (\se[\hat I])$
for samples of size 500 from each of the four populations when $a=1,2,3$ and 4.  \label{table6}}
\begin{tabular}{cccccc}
\toprule
 &\multicolumn{5}{c}{Gross Household Weekly Income (GWI)}\\[.2cm]
 \cline{2-6} &\multicolumn{2}{c}{1995}&&\multicolumn{2}{c}{2010}\\
$a$ &$\hat G$  & $\hat I$ &\qquad & $\hat G$ & $\hat I$   \\
1  & 0.519 (0.091) & 0.632 (0.012) &&  0.523 (0.057) & 0.626 (0.013)  \\
2  & 0.397 (0.014) & 0.604 (0.013) &&  0.465 (0.023) & 0.647 (0.013)  \\
3  & 0.408 (0.011) & 0.643 (0.013) &&  0.405 (0.014) & 0.622 (0.013)  \\
4  & 0.420 (0.015) & 0.627 (0.013) &&  0.438 (0.017) & 0.627 (0.013)  \\[.2cm]

 \toprule
&\multicolumn{5}{c}{Disposable Personal Weekly Income (DWI)}\\[.2cm]
  \cline{2-6}
    &\multicolumn{2}{c}{1995}&&\multicolumn{2}{c}{2010}\\
 $a$  &$\hat G$  & $\hat I$ &\qquad & $\hat G$ & $\hat I$   \\
1  &  0.360 (0.028) & 0.516 (0.013) && 0.389 (0.031) & 0.520 (0.013)    \\
2  &  0.287 (0.011) & 0.491 (0.013) && 0.419 (0.024) & 0.543 (0.013)    \\
3  &  0.287 (0.009) & 0.498 (0.012) && 0.346 (0.016) & 0.535 (0.013)    \\
4  &  0.308 (0.009) & 0.537 (0.013) && 0.339 (0.016) & 0.516 (0.013)    \\
\bottomrule
\end{tabular}
\end{footnotesize}
\end{center}
\end{table}

Now we are able to show the results of taking random samples of size $n=500$ from each of the four populations. For the case of $a=4$ density plots of these samples are shown in dashed lines in
Figure~\ref{fig:absdata}. In Table~\ref{table6} we can see that for $a=1$ or 2 the estimated
standard errors of $\hat G$ are larger than for $a=3$ and $a=4$, making comparisons between
results from different years 1995 and 2010 highly dependent on the unknown $a$,  which is difficult
to estimate.  On the other hand, for $\hat I$ the standard errors do not depend at all on these
choices of $a$.

\subsection{Numbers of visits to Doctors, by gender}\label{ex4}

 A major Health and Retirement study in the United States surveys adults every 2 years after they reach the
 age of 50.  The AHEAD cohort consisted of persons born in the United States before 1924 (and their spouses,
 regardless of age). The data includes observations
on the number of visits to doctors and 24 concomitant variables; it is analyzed by classical MLE  and robust regression methods in \cite[Sec.\;5.6]{HCCV-2009}. The data are found via the website
\begin{small}
\begin{verbatim}
http://http://www.unige.ch/gsem/rcs/members2/profs/eva-cantoni/books/
\end{verbatim}
\end{small}

\begin{figure}[h!t]
\begin{center}
\includegraphics[scale=.8]{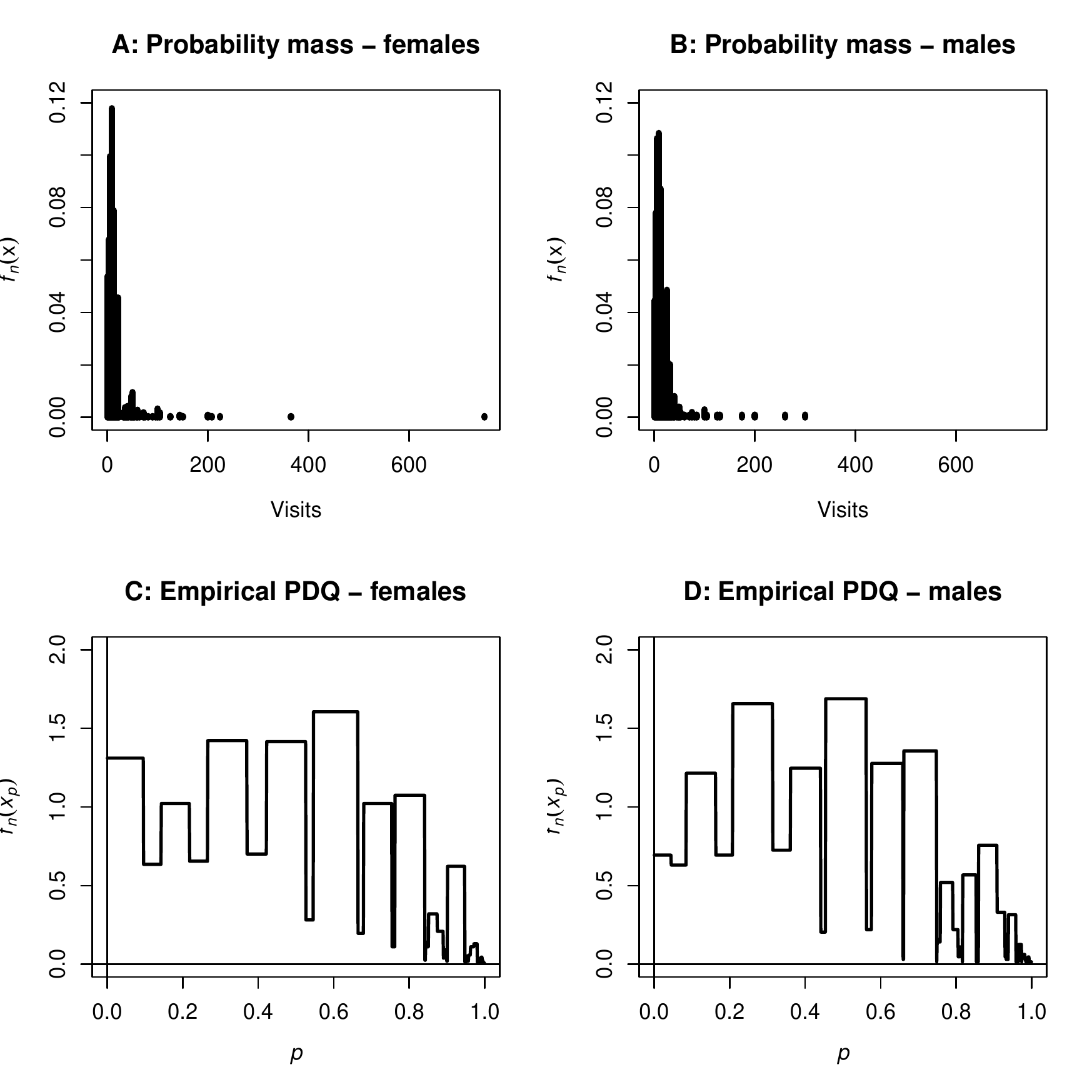}
\caption{\footnotesize \em Empiricial mass functions (Plots A and B) for the doctor visits data separated according to gender.  The normalized inverse of the estimated quantile density function for each is provided in Plots C and D.
\label{fig:visitsdata}}
\end{center}
\end{figure}

An empirical probability mass function for females and males is provided in Plots A and B of Figure \ref{fig:visitsdata} respectively.  It is immediately clear that the extrema for females are larger than that for males, suggesting increased variability in the female population.

\begin{table}[ht]
\begin{center}
\caption{{\bf Comparison of doctors visits by gender with $\hat G$ and $\hat I$.}\em
 Estimates of the inequality coefficients $G$ and $I$, including their standard errors in parentheses, and 95\%
confidence intervals.  M* refer the the largest value from the male data set having been removed.\label{table7}}
\begin{tabular}{lccccc}
\toprule
 gender     &  $\hat G$       &    $[L,U]_G$      &&  $\hat I$  &   $[L,U]_I$    \\[.3cm]
\cline{2-3}  \cline{5-6}
   F        &  0.520 (0.019)  & $[0.481, 0.559]$  &&  0.608 (0.007) & $[0.595, 0.622]$  \\[.2cm]
   M        &  0.487 (0.019)  & $[0.449, 0.525]$  &&  0.612 (0.010) & $[0.593, 0.631]$  \\[.2cm]
  F-M       &  0.033 (0.028)  & $[-0.021, 0.087]$  && $ -0.004 (0.012) $  & $[-0.027, 0.020]$ \\
  M*        &  0.476 (0.017)  & $[0.443, 0.510]$  &&  0.611 (0.010) & $[0.593, 0.630]$  \\[.2cm]
  F-M*       &  0.044 (0.026)  & $[-0.007, 0.095]$  && $ -0.003 (0.012) $  & $[-0.026, 0.020]$ \\
\bottomrule
\end{tabular}
\end{center}
\end{table}

In Table \ref{table7} we provide point and interval estimates for each of $G$ and $I$ for males and females and also estimates for the different between the two.  Both measures suggest a moderate degree of inequality and, interestingly, neither suggest that there is a difference in inequality between males and females.  In Plots C and D we provide the reciprocal of the empirical quantile density function for females and males.  The shapes of each are similar which further confirms that inequality should be approximately the same.  We can show with this data set that the estimator of $G$ is more heavily influenced by just one observation.  We removed the largest observation from the male data set and recalculated the intervals (these are denoted by M* in the table).  While the estimates for $I$ and the difference for $I$ between males and females are almost identical to those from the full data, we note a bigger change when using the Gini index.  In fact, the estimated interval for the difference between males and females comes very close to suggesting a significant difference and many would use this interval to conclude that there is indeed a difference.

\section{Summary  and further possible work}\label{sec:summary}

We proposed an inequality measure $I$ that depends only on the area between the symmetric ratio of
quantiles curve $\{(p,R(p)\}$ and the horizontal line at one.  The measure  $I$  has the simple interpretation as the average relative distance of a randomly chosen income from the lower half of incomes to its symmetric quantile.  This inequality measure is easy to estimate using distribution-free methods,
and is demonstratively resistant to outliers.  Despite its simplicity, in many cases it is a more effective measure than the Gini index which can be non-robust and more heavily biased, as shown in the simulations and examples. An R script for plotting the inequality curve $\{(p,R(p)\}$ and finding the estimates of $I$ and
its standard error, as well as confidence intervals for $I$ is included in supplementary on-line material.

In the Appendix we considered some examples illustrating cases where the inequality curve is convex, and
it would be of interest to find simple conditions on the underlying income distribution for which this is
the case.  Another research area of interest is to find a quantile-based measure of inequality for which
the measure applied to a mixture of income distributions is the same or related mixture of the inequality measures of the components, at least to a good approximation.  Finally it would be useful to show how
various factors  affect $I$, or some function thereof, in a regression setting.

\bibliography{srq}
\bibliographystyle{authordate4}

\clearpage
\newpage

\appendix

\begin{table}[ht]
\begin{center}
\caption{\footnotesize {\bf Examples of distributions $F(x)$ and associated functions.\ } \em In general, we denote $x_p=Q(p)=F^{-1}(p)$, but for the normal $F=\Phi $ with density $\varphi $, we write
$z_p=\Phi ^{-1}(p)$. The support of each $F$ is $(0,+\infty )$, except for the normal and Type I Pareto, the latter having
support on $[1,+\infty)$. \label{table8}}
\vspace{0.2cm}
\begin{footnotesize}
\begin{tabular}{lccccccc}
\toprule
                    &  $1-F(x)$            && $Q(p)$                              && $q(p)$ && $J(p)$             \\[.2cm]
\hline
 Exponential        &  $e^{-x} $           && $-\ln (1-p)$                        &&  $(1-p)^{-1} $
&& 1    \\[.3cm]
 Normal             &  $\Phi (-x)$         && $z_p$                               && $\frac {1}{\varphi (z_p)}$
&&  $z_p$  \\[.3cm]
 Lognormal          &  $ \Phi (-\ln (x))$  && $e^{z_p}$                           && $\frac {e^{z_p}}{\varphi(z_p)}$
&&  $ \frac {1+z_p}{e^{z_p}}$ \\[.3cm]
 Type I Pareto$(a)$ &  $ x^{-a} $          && $\frac{1}{(1-p)^{1/a}}$             &&$\frac{1}{a(1-p)^{1/a+1}}$
&&  $-(a+1)(1-p)^{1/a}$     \\[.3cm]
Type II Pareto$(a)$ &  $ (1+x)^{-a} $      &&$ \frac{1}{(1-p)^{1/a}}-1$           &&$\frac{1}{a(1-p)^{1/a+1}}$
&&  $-(a+1)(1-p)^{1/a}$   \\[.3cm]
 Weibull$(\beta)$   &  $ e^{-x^{\beta}} $  &&$[-\ln (1-p)]^{1/\beta}$ &&
 $\frac {\{-\ln (1-p)\}^{1/\beta -1}}{(1-p)\beta }$ && $\frac{ 1-\beta-\beta \ln(1-p)}{[-\ln (1-p)]^{1/\beta }}$ \\ \bottomrule
\end{tabular}
\end{footnotesize}
\end{center}
\end{table}

\section{Variance of the inequality index}

Firstly, using results from, for example, Chapter 7 of \cite{Das-2008}, as $n$ increases without bound  $\e (\hat x_p)\doteq x_p$ and
\begin{equation}\label{covxpxq}
n\,\cov (\widehat x_p,\widehat x_r)\doteq
  \begin{cases}
  p(1-r)q(p)q(r), &0 < p < r < 1\\
  r(1-p)q(p)q(r), &0 < r < p < 1
  \end{cases}.
\end{equation}

Now, for $0<p<r<1$, using the Delta method we have that
\begin{align*}
{\sigma}_{pr}\equiv n\,\cov [\widehat{R}(p),\widehat{R}(r)] \doteq &\frac{1}{x_{1-p/2}}\cdot\frac{1}{x_{1-r/2}}\Big[\cov(\widehat x_{p/2},\widehat x_{r/2})-\widehat{R}(r)\cov(\widehat x_{p/2},\widehat x_{1-r/2})\\
&-\widehat{R}(p)\cov(\widehat x_{1-p/2},\widehat x_{r/2})+\widehat{R}(p)\widehat{R}(r)\cov(\widehat x_{1-p/2},\widehat x_{1-r/2})\Big]
\end{align*}
so that, from \eqref{covxpxq} and noting that $p/2<r/2$, $p/2<1-r/2$, $1-p/2>r/2$ and $1-p/2>1-r/2$
\begin{align}
{\sigma}_{pr} \doteq \frac{1}{x_{1-p/2}}\cdot\frac{1}{x_{1-r/2}}\Big[&\frac{p}{2}\cdot\left(1-\frac{r}{2}\right)\left\{q\left(\frac{p}{2}\right)q\left(\frac{r}{2}\right)+\widehat{R}(p)\widehat{R}(r)q\left(1-\frac{p}{2}\right)q\left(1-\frac{r}{2}\right)\right\}\nonumber\\
&-\frac{p}{2}\cdot\frac{r}{2}\left\{\widehat{R}(r)q\left(\frac{p}{2}\right)q\left(1-\frac{r}{2}\right)+\widehat{R}(p)q\left(1-\frac{p}{2}\right)q\left(\frac{r}{2}\right)\right\}\Big].\label{covRpRq}
\end{align}

Finally, we have that
\begin{equation}
\text{Var}\left[\widehat{I}(J)\right]=\frac{1}{n}\cdot\frac{1}{J^2}\left[\sum^J_{j=1}\widehat{\sigma}^2_{p_j}+2\sum^{j-1}_{i=1}\sum^J_{j = 2}\widehat{\sigma}_{p_ip_j}\right]\label{VarI}
\end{equation}
where $p_j=(j - 1/2)/J$ for $j=1,\ldots,J$, $\widehat{\sigma}_{p_iq_j}$ and $\widehat{\sigma}_{p_j}$ are the estimates to ${\sigma}_{pr}$ and ${\sigma}_{p}$ in \eqref{covRpRq} and \eqref{ASVR}.  The estimates are obtain be replacing the population quantiles with the respective estimates and quantile density function, $q$, with an appropriate estimate.

\section{Convexity of R}\label{sec:convex}

In this section we restrict attention to $F\in \F ''$, defined by
\begin{equation*}
   \F ''=\{F\in \F ':\ f'=F'' \text { exists.} \}
\end{equation*}
This guarantees the existence of the quantile density $q(p)=Q'(p)$ as well as its derivative $ q'(p) = -\;f'(x_p)/f^3(x_p)=J(p)q^2(p)$.
 For possible further use we note that the {\em score function} for a location-scale family is defined for each $p$ by $J(p)=-f'(x_p)/f(x_p)$; it arises in nonparametric statistics \cite{hajek-1967}.  \cite{par-1979} notes that  $J(p)=-\frac {d}{dp}\; f(Q(p))$.
Further, the score function for a {\em scale} family is defined for each $p$ by $K(p)=-1+Q(p)J(p).$ Some examples
of these functions are collected in Table~\ref{table8}. Note that $K(p)=-a-2$ is free of $p$ for the Type I Pareto distribution.

We now seek restrictions on the family $\F ''$ for which the quantile ratio curves are convex.
To examine the convexity of $R$ we need to find an expression for $R''$, and to find
 the derivatives of $R$ it is convenient to introduce
$H_Q(p)=\frac {\partial \;\ln (Q(p))}{\partial p}=q(p)/Q(p)$ for $0<p<1$.
Then
\begin{equation}\label{Hpr}
    H'_Q(p)=\frac{q'(p)}{Q(p)}-\frac{q^2(p)}{Q^2(p)}=K(p)H^2_Q(p)~.
\end{equation}
Therefore
 \begin{eqnarray}\label{derivR} \nonumber
 R(p) &=& \frac {Q \left (\frac{p}{2}\right )}{Q\left (1-\frac{p}{2}\right )} \\ \nonumber
R'(p) &=& \frac {q \left (\frac{p}{2}\right )}{2Q\left (1-\frac{p}{2}\right )}
     +\frac {q\left (1-\frac{p}{2}\right )Q\left (\frac{p}{2}\right )}{2Q^2\left (1-\frac{p}{2}\right )}
 \\ \nonumber
&=& \frac{R(p)}{2}\;\left \{H_Q\left (\frac{p}{2}\right )+H_Q\left (1-\frac{p}{2}\right )\right \}\\ \nonumber
 R''(p)&=& \frac{R'(p)}{2}\;\left \{H_Q\left (\frac{p}{2}\right )+H_Q\left (1-\frac{p}{2}\right )\right \}+\frac{R(p)}{4}\;\left \{H_Q'\left (\frac{p}{2}\right )-H_Q'\left (1-\frac{p}{2}\right )\right \}\\ \nonumber
  &= &  \frac{R(p)}{4}\;\left [\left \{H_Q\left (\frac{p}{2}\right )+H_Q\left (1-\frac{p}{2}\right )\right \}^2+\left \{H_Q'\left (\frac{p}{2}\right )-H_Q'\left (1-\frac{p}{2}\right )\right \}\right ]            ~.
\end{eqnarray}
Using (\ref{Hpr}), the term in square brackets is positive if and only if
\[ \left \{H_Q\left (\frac{p}{2}\right )+H_Q\left (1-\frac{p}{2}\right )\right \}^2+ K\left (\frac{p}{2}\right )
H^2_Q\left (\frac{p}{2}\right ) - K\left (1-\frac{p}{2}\right )H^2_Q\left (1-\frac{p}{2}\right ) >0 ;\]
that is, if and only if $t(p)>0$ for all $0<p<1$, where $t$ is defined by:
\begin{equation}\label{t}
    t(p)\equiv H_Q^2\left (\frac{p}{2}\right )\left \{1+K\left (\frac{p}{2}\right )\right \}  +
H_Q^2\left (1-\frac{p}{2}\right ) \left \{1-K\left (1-\frac{p}{2}\right )\right \}
+2H_Q\left (\frac{p}{2}\right )H_Q\left (1-\frac{p}{2}\right )~.
\end{equation}

We will now determine the convexity of some important income distributions by determining whether $t(p)$ versus $p$  is positive.

\subsection{Type I Pareto}

Using Table~\ref{table8}, $H_Q(p)=\{a(1-p)\}^{-1}$ and $K(p)=-1+Q(p)J(p)\equiv -a-2.$   Hence $t$ defined by (\ref{t})
equals:
\begin{eqnarray*}
t(p) &=&  -(a+1)  H_Q^2\left (\frac{p}{2}\right )+(3+a)H_Q^2\left (1-\frac{p}{2}\right )
+2H_Q\left (\frac{p}{2}\right )H_Q\left (1-\frac{p}{2}\right )\\
   &=& -\frac {4(a+1)}{a^2(2-p)^2}+\frac {4(a+3)}{a^2p^2}+\frac {8}{a^2p(2-p)}\\
   &=& \frac{4}{a^2p^2(2-p)^2}\left \{ -(a+1)p^2+(a+3)(2-p)^2+2p(2-p)\right \} \\
   &=& \frac{16}{a^2p^2(2-p)^2}\left [(a+3)(1-p)+p\right ]
\end{eqnarray*}
which is positive for all $a>0$ and $0<p<1.$

\subsection{Type II Pareto}
\begin{figure}[h!t]
\begin{center}
\includegraphics[scale=.6]{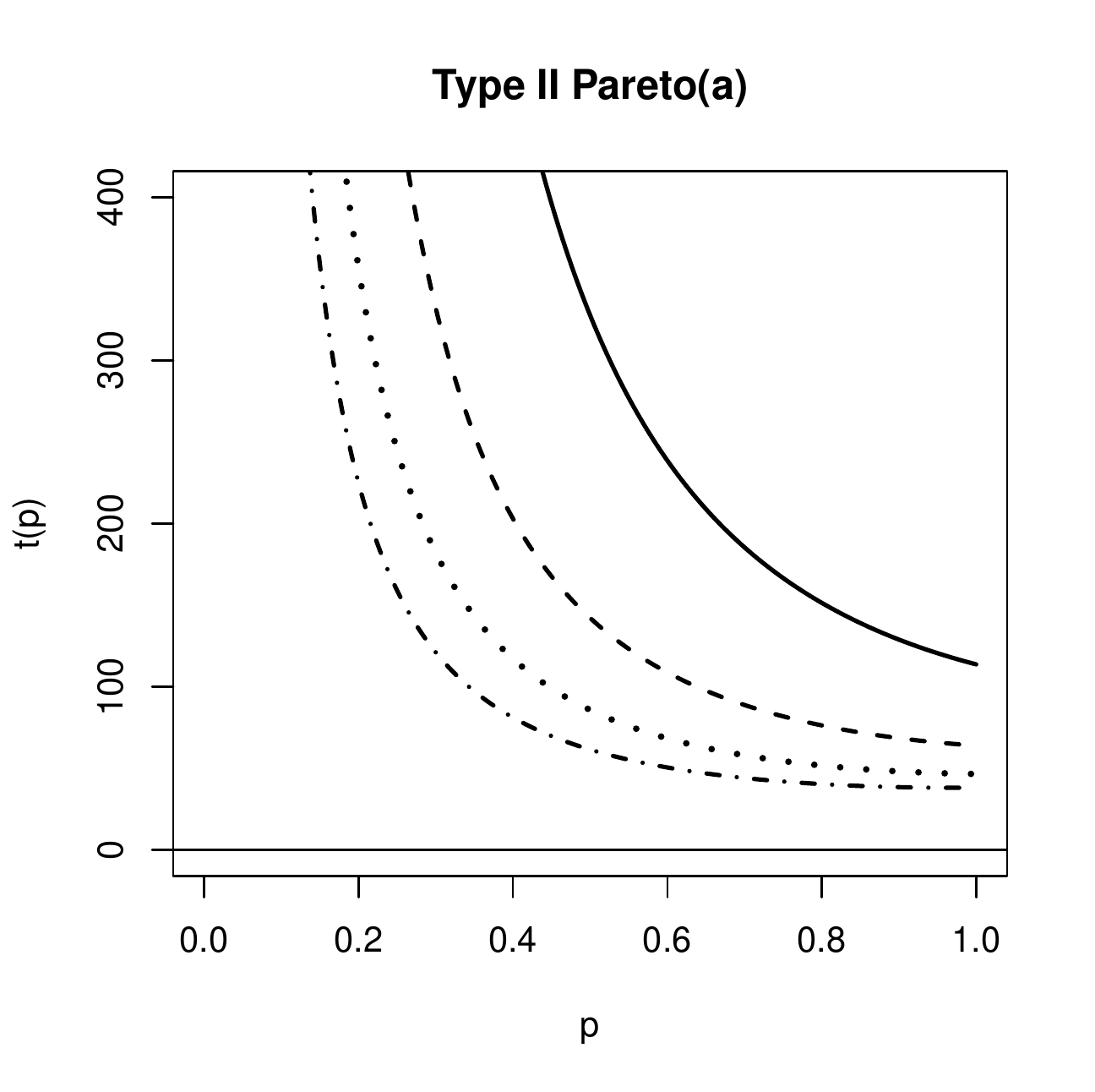}
\includegraphics[scale=.6]{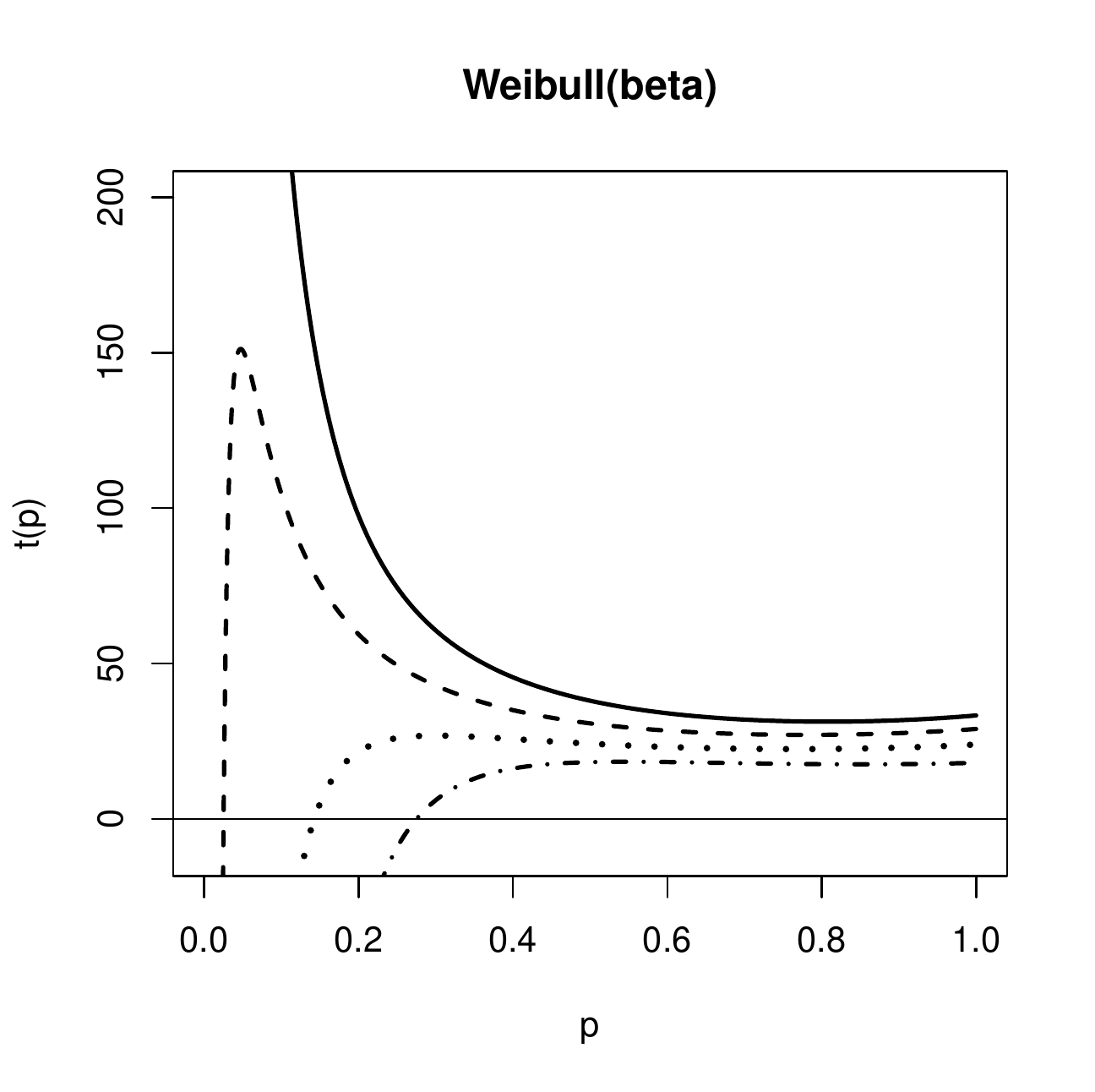}
\caption{\footnotesize \em Plot of $t(p)$ versus $p$ for the Type II Pareto($a$) model and for the Weibull($\beta$) model. For the Type II Pareto, the solid line is for
 $a=0.5$, the dashed line for $a=1$, and dotted line $a=2$ and and the dot-dashed line $a=5$. For the Weibull, the solid line is for
 $\beta=1$, the dashed line for $\beta =1.1$, the dotted line for $\beta =1.25$ and and the dot-dashed line $\beta =1.5$. \label{fig4}}
\end{center}
\end{figure}

The function $t(p)$ is messier to compute for the Type II Pareto model, but a plot of $t(p)$ versus $p$ for various $a$ is shown in Figure~\ref{fig4}. These and other plots convince us that $R$ is convex for all values of $a>0.$

\subsection{Weibull}

The function $t(p)$ is also messy to compute for the Weibull model, but a plot of $t(p)$ versus $p$ for various $a$ is shown in Figure~\ref{fig4}. These and other plots suggest that $t(p)>0$ for all $0<p<1$ (and hence the ratio $R$ is convex) for
all $\beta <1.04$. For $\beta >1.05$ they are negative for small $p$, and hence not convex; some examples are plotted
in Figure~\ref{fig1}.

What the above examples show is that convexity of the ratio of symmetric quantiles $R(p)=x_{p/2}/x_{1-p/2}$ is not a common attribute for $F\in \F$, although it does hold for Type I and II Pareto models,  Weibull models with small $\beta $ (including the exponential), and that it holds for the log-normal model except when $p<0.045,$ where it is nearly linear.
This lack of convexity might be considered a disadvantage relative to Lorenz curves, which are always convex when they are
defined, but $R$ is not only defined for all $F\in \F$, it has a greater range of values and can be estimated by distribution-free methods.

\setlength{\tabcolsep}{3pt}
\renewcommand{\arraystretch}{1}
\pagestyle{empty}
\begin{landscape}

\section{Bias and standard error of $\hat G$ and $\hat I$}\label{sec:appbias}

\begin{table}[h!]
\centering
\begin{footnotesize}
\begin{tabular}{lrraaaarrrraaaarrrraaaa}
  \toprule
 & & & \multicolumn{4}{c}{$n=50$} & \multicolumn{4}{c}{$n=100$} & \multicolumn{4}{c}{$n=200$} & \multicolumn{4}{c}{$n=500$} & \multicolumn{4}{c}{$n=1000$}\\
\# & $I$ & $G$ & $b_I$ & $b_G$ & $s_I$ & $s_g$ & $b_I$ & $b_G$ & $s_I$ & $s_g$ & $b_I$ & $b_G$ & $s_I$ & $s_g$ & $b_I$ & $b_G$ & $s_I$ & $s_g$ & $b_I$ & $b_G$ & $s_I$ & $s_g$ \\ \hline
1& 0.664& 0.520& -0.013& -0.018& 0.040& 0.055& -0.006& -0.009& 0.027& 0.039& -0.004& -0.004& 0.020& 0.028& -0.001& -0.002& 0.012& 0.018& -0.001& -0.001& 0.008& 0.014\\
2& 0.915& 0.489& -0.026& -0.001& 0.061& 0.062& -0.015& -0.002& 0.043& 0.045& -0.007&  0.000& 0.029& 0.032& -0.003&  0.001& 0.018& 0.020& -0.002&  0.000& 0.013& 0.014\\
3& 0.727& 0.405& -0.014& -0.003& 0.056& 0.045& -0.009& -0.003& 0.037& 0.031& -0.005& -0.002& 0.027& 0.022& -0.001&  0.000& 0.017& 0.014& -0.001& -0.001& 0.012& 0.010\\
4& 0.614& 0.333& -0.015& -0.004& 0.050& 0.033& -0.009& -0.003& 0.034& 0.024& -0.003& -0.001& 0.025& 0.017& -0.001& -0.001& 0.016& 0.011&  0.000&  0.001& 0.011& 0.008\\
5& 0.280& 0.124& -0.011& -0.002& 0.025& 0.012& -0.005& -0.001& 0.018& 0.009& -0.003& -0.001& 0.013& 0.006& -0.002&  0.000& 0.008& 0.004&  0.000&  0.000& 0.006& 0.003\\
6& 0.801& 0.637& -0.015& -0.015& 0.039& 0.045& -0.007& -0.007& 0.028& 0.032& -0.002& -0.002& 0.020& 0.022& -0.001& -0.002& 0.012& 0.014& -0.001&  0.000& 0.008& 0.010\\
7& 0.595& 0.375& -0.012& -0.007& 0.041& 0.034& -0.008& -0.005& 0.028& 0.023& -0.004& -0.002& 0.020& 0.017& -0.002& -0.001& 0.013& 0.011& -0.001&  0.000& 0.009& 0.007\\
8& 0.333& 0.158& -0.012& -0.004& 0.029& 0.016& -0.006& -0.002& 0.019& 0.011& -0.003& -0.001& 0.014& 0.008& -0.002&  0.000& 0.009& 0.005& -0.001&  0.000& 0.006& 0.003\\
9& 0.773& 0.997& -0.011& -0.217& 0.037& 0.092& -0.005& -0.186& 0.028& 0.082& -0.003& -0.161& 0.019& 0.069&  0.000& -0.139& 0.012& 0.056& -0.001& -0.125& 0.008& 0.048\\
10& 0.740& 0.667& -0.013& -0.036& 0.039& 0.074& -0.007& -0.022& 0.028& 0.060& -0.004& -0.012& 0.020& 0.045& -0.001& -0.004& 0.012& 0.035& -0.001& -0.004& 0.009& 0.026\\
11& 0.702& 0.503& -0.014& -0.012& 0.042& 0.042& -0.007& -0.005& 0.029& 0.029& -0.002& -0.002& 0.020& 0.021& -0.001& -0.001& 0.013& 0.013& -0.001&  0.000& 0.009& 0.009\\
12& 0.835& 0.750& -0.013& -0.022& 0.035& 0.050& -0.006& -0.011& 0.025& 0.038& -0.002& -0.005& 0.017& 0.026& -0.001& -0.002& 0.011& 0.017&  0.000& -0.001& 0.008& 0.012\\
13& 0.702& 0.500& -0.014& -0.012& 0.040& 0.041& -0.006& -0.005& 0.029& 0.029& -0.004& -0.003& 0.021& 0.021& -0.001& -0.001& 0.013& 0.013&  0.000&  0.000& 0.009& 0.009\\
14& 0.523& 0.293& -0.014& -0.006& 0.041& 0.028& -0.008& -0.004& 0.027& 0.019& -0.004& -0.002& 0.021& 0.014& -0.001&  0.000& 0.013& 0.009&  0.000&  0.000& 0.009& 0.006\\
15& 0.167& 0.067& -0.008& -0.001& 0.018& 0.008& -0.004& -0.001& 0.013& 0.005& -0.003& -0.001& 0.009& 0.004& -0.001&  0.000& 0.006& 0.002& -0.001&  0.000& 0.004& 0.002\\
16& 0.543& 0.534& -0.015& -0.072& 0.045& 0.117& -0.006& -0.044& 0.031& 0.096& -0.004& -0.035& 0.022& 0.075& -0.001& -0.019& 0.015& 0.068& -0.002& -0.017& 0.010& 0.050\\
\bottomrule
\end{tabular}
\end{footnotesize}
\caption{Simulated bias and standard deviation (from 1,000 trials) for varying distributions and sample sizes.  The simulated bias for $\widehat{I}$ and $\widehat{G}$ are labeled $b_I$ and $b_G$ respectively.  Standard deviations are denoted $s_I$ and $s_G$}\label{table9}
\end{table}
\end{landscape}

\end{document}